\documentclass{aa}  

\usepackage{graphicx}
\usepackage{txfonts}
\usepackage{xspace}
\usepackage{rotating}
\usepackage{pdflscape}
\usepackage{longtable}
\usepackage{booktabs}
\usepackage{color}
\usepackage{url}
\usepackage{hyperref}

\usepackage{xcolor}
\hypersetup{
  colorlinks   = true, 
  urlcolor     = blue, 
  linkcolor    = blue, 
  citecolor   = blue 
}
\usepackage{siunitx}
\usepackage{rotating}

\usepackage[flushleft]{threeparttable}
\usepackage{graphicx,epstopdf}
\DeclareGraphicsExtensions{.ps}
\def\Msun{M_\odot}

\graphicspath{{./}{figures/}}
\begin{document}

   \title{Explanations for the two-component spectral energy distributions of gravitationally lensed stars at high redshifts}

   \author{Armin Nabizadeh
          \inst{1}
          \and
          Erik Zackrisson
          \inst{1,2}
          \and
          Emma Lundqvist
          \inst{1}
          \and
          Massimo Ricotti
          \inst{3}
          \and
          Seyong Park
          \inst{3}
          \and
          Brian Welch
          \inst{3,4}
          \and
          Jose M. Diego
          \inst{5}
          }

   \institute{Observational Astrophysics, Department of Physics and Astronomy, Uppsala University, Box 516, SE-751 20 Uppsala, Sweden\\
              \email{armin.nabizade@gmail.com}
    \and
         Swedish Collegium for Advanced Study, Linneanum, Thunbergsv\"a{}gen 2
SE-752 38 Uppsala, Sweden
    \and
    Department of Astronomy, University of Maryland, College Park, MD 20742, USA
    \and
     Observational Cosmology Lab, NASA Goddard Space Flight Center, Greenbelt, MD 20771, USA
    \and
    Instituto de F\'isica de Cantabria (CSIC-UC). Avda. Los Castros s/n. 39005 Santander, Spain
             }

\titlerunning{Two-component SEDs of lensed stars}
\authorrunning{Nabizadeh et al.}

   \date{}

\abstract{Observations of gravitationally lensed, high-mass stars at redshifts $\gtrsim1$ occasionally reveal spectral energy distributions that contain two components with different effective temperatures. Given that two separate stars are involved, it suggests that both stars have simultaneously reached very high magnification, as expected for two stars in a binary system close to the caustic curve of the foreground galaxy-cluster lens. The inferred effective temperatures and luminosities of these stars are, however, difficult to reconcile with known binaries, or even with isolated stars of the same age. Here, we explore three alternative explanations for these cases: circumstellar dust around the cooler of the two stars; age differences of a few Myr among stars in the same star cluster, and a scenario in which the stars originate in two separate star clusters of different age along the lensing caustic. While all of these scenarios are deemed plausible in principle, dust solutions would require more circumstellar extinction than seen in local observations of the relevant super/hypergiant stars. Hence, we argue that age differences between the two stars are the most likely scenario, given the current data.}


   \maketitle
%

\section{Introduction}
In recent years, observations with the Hubble Space Telescope (HST) and the James Webb Space Telescope (JWST) of galaxy-cluster fields have enabled the detection of dozens of gravitationally lensed stars beyond the local Universe \citep{Kelly18,Chen19,Kaurov19,Welch22a,Chen22,Kelly22,Diego23a,Meena23,Yan23,Diego23b,Fudamoto24}, with the highest-redshift example so far being the star Earendel at $z\approx 6$ \citep{Welch22a,Welch22b}. In some cases, the spectral energy distributions (SEDs) of these objects indicate the presence of two components with different effective temperatures ($T_\mathrm{eff}$), which has been interpreted as the blended light from two separate high-mass stars in the same star cluster or as two stars from a high-mass binary system \citep{Welch22b,Diego23b,Furtak24}. However, the ratios of bolometric luminosities between the hot and cool components (possibly a blue supergiant and a yellow/red super- or hypergiant) inferred from the SED fits, are puzzling. In the two-component fits to lensed stars presented by \citet{Welch22b}, \citet{Diego23b} and \citet{Furtak24} the best-fitting bolometric luminosity ratios are $L_\mathrm{hot}/L_\mathrm{cool}\gtrsim 2$, which in the context of single-star evolution would be unexpected for two stars of the same age. A star with a higher initial mass is generally expected to evolve from high to low $T_\mathrm{eff}$ at a faster pace and at a higher luminosity than a star with a lower initial mass. The latter is therefore expected to linger in a higher-$T_\mathrm{eff}$, lower-$L$ state as the higher-mass star reaches its low-$T_\mathrm{eff}$, high-$L$ end states, resulting in a bolometric luminosity ratio $L_\mathrm{hot}/L_\mathrm{cool}\lesssim 1$ for two stars of the same age and metallicity.

The binary-star hypothesis may seem plausible, given that most high-mass stars are in binaries systems \citep[e.g.,][]{Sana12}, and at redshift $z\gtrsim 1$, only high-mass stars ($\gtrsim 5$--10 $M_\odot$) are likely to be rendered sufficiently bright by lensing to be detectable \citep{2022MNRAS.514.2545M}. Binary evolution can give rise to more complicated evolutionary pathways \citep[for a recent review, see][]{Marchant23}, but in observed samples of red supergiants in binary systems \citep{Patrick22}, cases where the hotter companion has a higher luminosity are very rare (in the sample of 88 red supergiant binaries in \citealt{Patrick22}, all obey $L_\mathrm{hot}\lesssim L_\mathrm{cold}$). 

In this work, we explore three alternative explanations for the luminosity anomaly observed in the two-component SEDs of lensed stars: (I) the possibility that the $L_\mathrm{hot}/L_\mathrm{cool}$ ratio has been overestimated due to the presence of circumstellar dust around the low-$T_\mathrm{eff}$ star in a same-age system; (II) the possibility that two stars that differ in age by a few million years may co-exist in the same star cluster, and (III) that the two stars are located in different star clusters, with more disparate ages, but still appear as an unresolved pair when magnified. Section ~\ref{Section: Puzzle} outlines the puzzle of the $L_\mathrm{hot}/L_\mathrm{cool}$ luminosity ratios in lensed stars. Sections ~\ref{Section: Circumstellar dust}, \ref{Section: Same star cluster}, and \ref{Section: Different star clusters} explore scenarios I, II, and III, respectively. Section ~\ref{Section: Discussion} discusses and summarizes our findings and results.


\section{The puzzling two-component SEDs of lensed stars}
\label{Section: Puzzle}
The current examples of lensed stars that may require two-component SED fits include Mothra at $z\approx 2$ \citep{Diego23b}, Earendel at $z\approx 6$ \citep{Welch22a,Welch22b} and MACS0647-star-1 at $z\approx 5$ \citep{Meena23,Furtak24}. In our view, Mothra provides the most robust case, because this object shows variability in the red component but not in the blue, which clearly demonstrates that (at least) two separate light sources must be contributing to the overall SED. The SED of Earendel reported by \citet{Welch22b} is challenging to explain with the SED of a single star. However, it should be noted that all attempts to derive the photometry did not produce SEDs that would necessarily require such a solution \citep[see Fig. 6 in][]{Welch22b}. Furthermore, it is not clear whether the noisy JWST cycle-1 NIRSpec/prism spectrum of Earendel confirms the photometric SED (Welch 2024, in prep.). \citet{Furtak24} demonstrate that the JWST/NIRSpec spectrum of MACS0647-star-1 can be explained either as a two-component SED fit or a single, dust-reddened star. The possibility of it being an extreme star cluster is also suggested, although this explanation is disfavoured by the lensing situation.

   \begin{figure}
   \centering
   \includegraphics[width=\columnwidth]{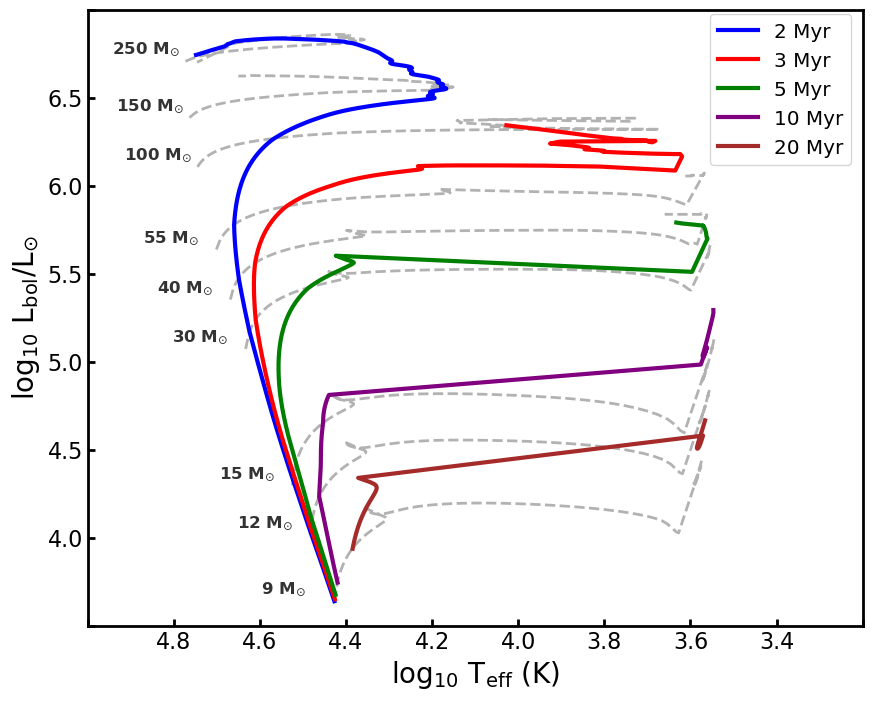}
      \caption{Evolutionary tracks for various ZAMS masses obtained from the BoOST SMC \citep[gray dashed lines;][]{BoOst-tracks2022} stellar evolutionary tracks, together with the corresponding isochrones interpolated for selected ages (colored solid lines). In this set of stellar evolutionary tracks, cases where two stars along the same isochrone may exhibit $T_\mathrm{hot}>12000$ K, $T_\mathrm{cool}$<8000 K, and $L_\mathrm{hot}/L_\mathrm{cool}\geq 2$ arise only at ages $\lesssim 3$ Myr, where the isochrones curve back towards high $T_\mathrm{eff}$ (red and blue lines) due to ``blue loops'' along the $M_\mathrm{ZAMS}> 55 \ M_\odot $ tracks.}
         \label{fig:isochrones}
   \end{figure}

What the SEDs of these lensed objects share in common is significant flux shortward of the Balmer break, with the break itself being very prominent for Earendel and MACS0647-star-1, and less so for Mothra. However, they exhibit a relatively flat continuum longward of the break. The former is, broadly speaking, expected for a star with $T_\mathrm{eff}>10000$ K (the blue continuum slope of Earendel shortward of the Balmer break moreover requires $T_\mathrm{eff}>20000$ K), whereas the latter is expected for a  $T_\mathrm{eff}<8000$ K star.  Since the $T_\mathrm{hot}$,  $T_\mathrm{cool}$ and $L_\mathrm{hot}/L_\mathrm{cool}$ solutions of these objects exhibit substantial degeneracies, we will assume throughout this paper that acceptable solutions have $T_\mathrm{hot}>12000$ K, $T_\mathrm{cool}$<8000 K, and $L_\mathrm{hot}/L_\mathrm{cool}\geq 2$.

Given typical stellar evolutionary tracks for single stars, solutions that match these criteria require very finely tuned ages and masses if one needs the two stars to have formed at the same time. This is exemplified using SMC-metallicity tracks from \citet{Szecsi22}, and isochrones derived from these, in Fig~\ref{fig:isochrones}. For this set of tracks, solutions are possible in cases where the more massive, higher-$L$ star evolves back towards high $T_\mathrm{eff}$ along a ``blue loop'' after a brief time in a lower-$T_\mathrm{eff}$ state. If a lower-mass, lower-luminosity star simultaneously happens to have evolved to low $T_\mathrm{eff}$, the required $T_\mathrm{hot}$, $T_\mathrm{cool}$, and $L_\mathrm{hot}/L_\mathrm{cool}$ conditions can in principle be met.

\begin{figure}
\centering
\includegraphics[width=\columnwidth]{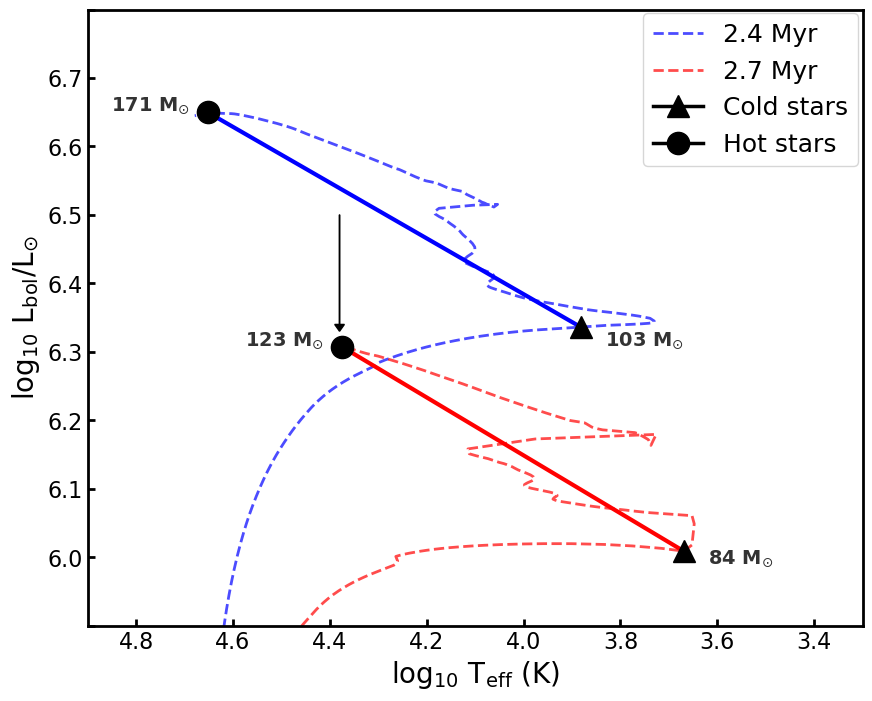}
  \caption{Examples of BoOST SMC Isochrones for which the  
$T_\mathrm{hot}>12000$ K, $T_\mathrm{cool}$<8000 K, and $L_\mathrm{hot}/L_\mathrm{cool}\geq 2$ conditions for lensed stars with two-component SEDs can be met. The isochrones are depicted with the blue and red dashed lines, obtained for ages 2.4 and 2.7, respectively. The blue and red solid lines connect stars with the same ages and different ZAMS masses that meet the requirements For clarity, the red isochrone has been shifted downward, as indicated by the black arrow.
          }
     \label{fig:solutions_same_age}
\end{figure}

By investigating isochrones between 1 and 27 Myr, separated by 0.1 Myr, we identified solutions for four isochrones at ages 2.4, 2.5, 2.6, and 2.7 Myr where the cold and hot components exhibit $L_\mathrm{hot}/L_\mathrm{cool}\geq 2$. Two such solutions are shown in Fig.~\ref{fig:solutions_same_age}. These solutions require a cold component with a Zero Age Main-Sequence (ZAMS) mass $\sim 83$--105$\,\Msun$  and a hot component with a ZAMS mass in range $\sim 121$--176$\,\Msun$ which corresponds to a narrow mass ratio of $R_{\rm M} = M_{\rm ZAMS, cold}/M_{\rm ZAMS, hot} = 0.58$--0.68. The limited number of isochrones and the narrow range of ZAMS masses yielding acceptable solutions indicate that such cases must be exceedingly rare (solutions in only 4 isochrones out of 270 studied matched the criteria). This demonstrates why it generally appears unlikely for two massive stars (ZAMS mass $M\geq 9\,\Msun$) of the same age to account for the observed two-component SEDs of lensed stars under the assumption of single-star evolution, whether they are two genuine single stars from a star cluster where all stars have the same age or members of a non-interacting binary system.

\begin{figure*}
    \centering
    \includegraphics[width=\columnwidth]{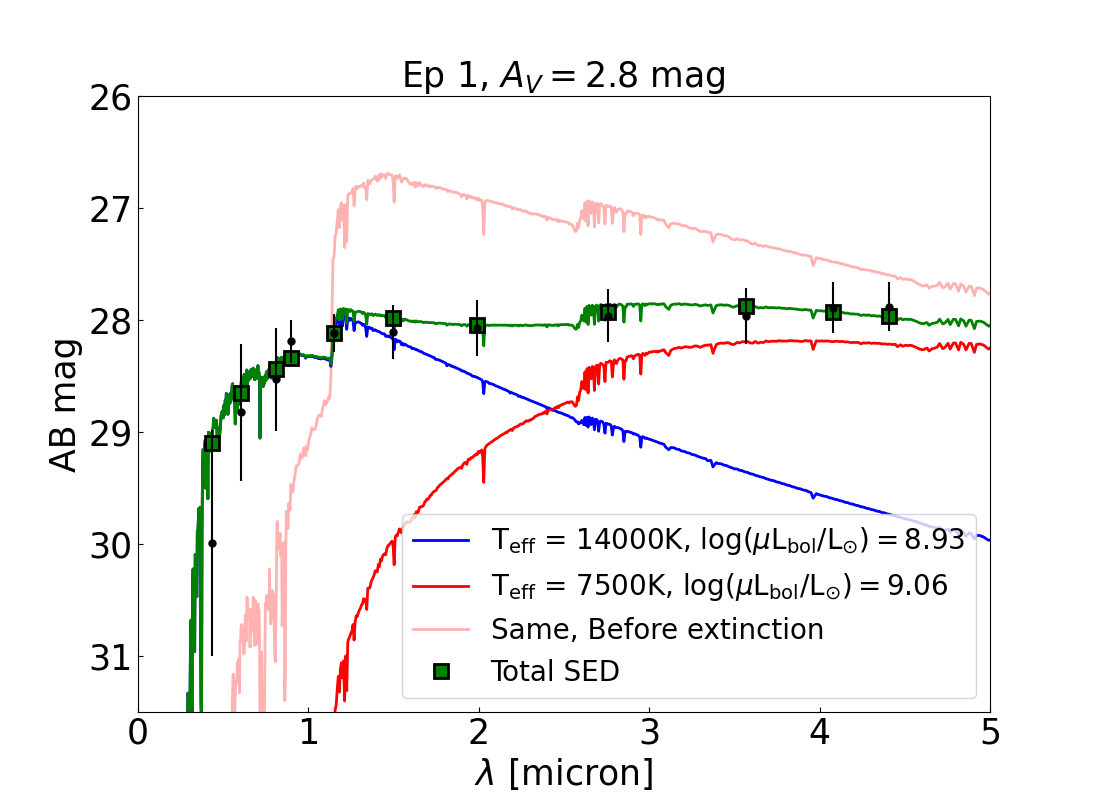}
    \includegraphics[width=\columnwidth]{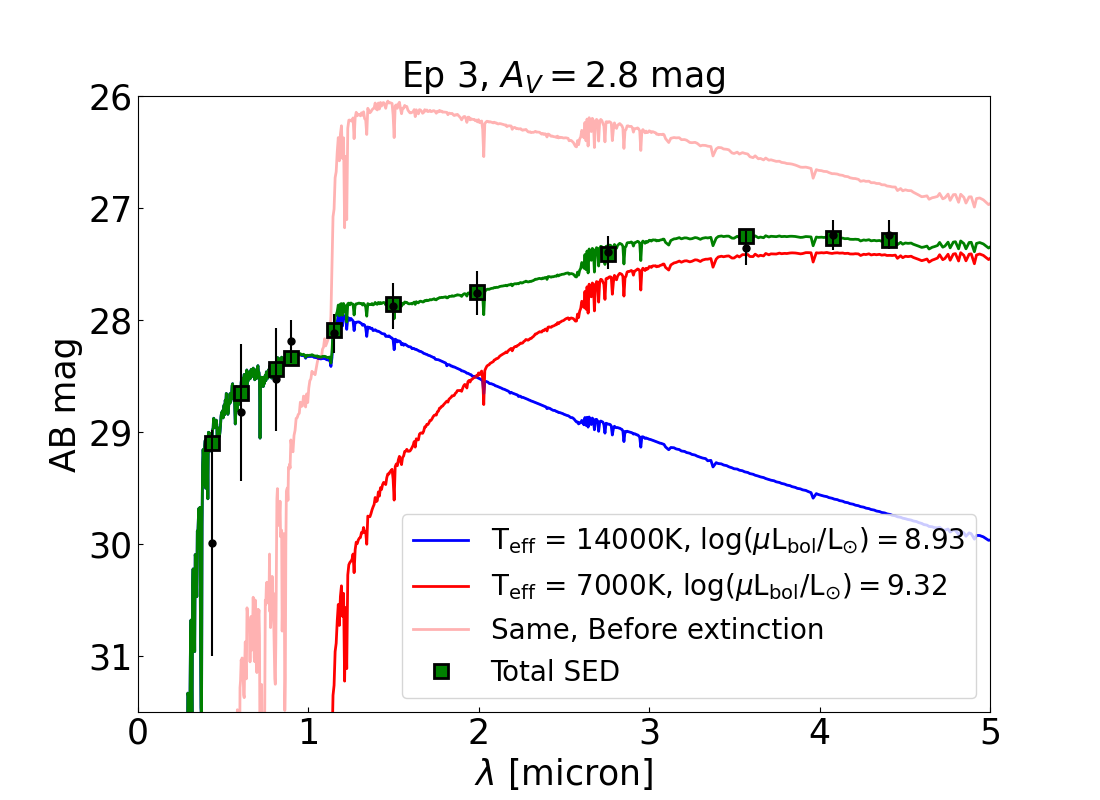}
    
    \caption{The effect of including circumstellar dust around the cooler component in the two-star solution for the observed Mothra SED. Two fits have been conducted for the SEDs, corresponding to epoch 1 (left) and epoch 3 (right), considering the variability of the cooler component. In the fits presented, the effective temperature and luminosity of the cooler component are both allowed to vary between the epochs. As seen, significant circumstellar extinction (here $A_V=2.8$ mag) allows for good fits to the SEDs of both epochs while keeping $L_\mathrm{hot}/L_\mathrm{cool} <1$.}
\label{fig:extictions}
\end{figure*}

\section{Scenario I: Circumstellar dust around the low-$T_\mathrm{eff}$ component}
\label{Section: Circumstellar dust}

A potential explanation for the anomalous luminosity ratios observed between the two SED components in lensed stars such as Mothra and Earendel is that the constituent stars have the same age, but the low-$T_\mathrm{eff}$ star experiences more extinction compared to the high-$T_\mathrm{eff}$ star. If this scenario is overlooked during the fitting procedure, where observed SEDs are fitted to models assuming equal or no extinction for both stars, this results in an underestimation of the luminosity of the low-$T_\mathrm{eff}$ object.

In a scenario where the two stars are members of the same star cluster, regardless of whether they are also part of the same binary system, they may potentially experience comparable levels of extinction from both the interstellar medium within the cluster and the ambient interstellar medium of the host galaxy. However, if the stars are not in the same binary system, and one of the stars has been actively producing dust in its immediate surroundings, the SED of that star may experience an additional, potentially dominant extinction component, which dims its apparent brightness. Both red supergiants and yellow supergiants are known to be surrounded by circumstellar dust, sometimes resulting in several magnitudes of visual extinction \citep[e.g.][]{Massey05,Gordon16}. One example of a star where dust extinction is blamed for dimming is R Coronae Borealis, where varying dust extinction causes an apparent luminosity change of over 8 magnitudes \citep{OKeefe39}. Here, we investigate the extent to which solutions involving circumstellar dust around the low-$T_\mathrm{eff}$ component can explain the properties of Mothra.

As discussed in \citet{Diego23b}, the long-wavelength part of the Mothra SED displays significant variability for observed wavelengths $\gtrsim 1.5\ \mu$m, whereas the short-wavelength part does not. The observed variability coupled with the lower redshift of Mothra, enables the existing JWST/NIRCam observations to probe the SED at rest-frame near-infrared wavelengths. This provides stronger constraints to be placed on the nature of the low-$T_\mathrm{eff}$ component compared to Earendel. In \citet{Yan23}, the JWST/NIRCam SEDs for Mothra are presented at three separate epochs, in which the long-wavelength component displays the lowest brightness in epoch 1 and the highest in epoch 3. These epochs are separated by $\approx 126$ days in the observed frame, corresponding to a rest-frame variability timescale of $\approx 41$ days at the redshift of Mothra, under the assumption that the variability is entirely intrinsic to the source and not affected by changes in microlensing magnification between the two components.

In Fig.~\ref{fig:extictions}, we demonstrate that by combining the HST observation of \citet{Diego23b} with the JWST observation of epochs 1 and 3 from \citet{Yan23}, it is possible to find acceptable SED fits that include circumstellar dust around the low-$T_\mathrm{eff}$ star and results in a non-anomalous intrinsic luminosity ratio of $L_\mathrm{hot}/L_\mathrm{cool}\approx 0.74$. 
This SED solution is, not unique, as it is possible to match both the overall SED and its change over time with models that allow variations in circumstellar dust extinction, luminosity, effective temperature, or some combination thereof, between the epochs. The solution shown in Fig.~\ref{fig:solutions_different_ages} assumes the circumstellar extinction remains constant between the observed epochs, and follows a Milky Way reddening law ($R_V=A_V\times E(B-V)=3.1$) as parameterized by \citet{Li08} in the Mothra rest frame. 

Including circumstellar dust in the fit has the effect of raising the intrinsic luminosity of the low-$T_\mathrm{eff}$ component. Additionally, assuming a non-gray extinction law (i.e., one that produces reddening) also increases the effective temperature ($T_\mathrm{eff}$) of the best-fitting low-$T_\mathrm{eff}$ component compared to a fit that assumes negligible extinction. In the extinction-free fit presented by \citet{Diego23b}, but with otherwise identical SED models, the effective temperature of the low-$T_\mathrm{eff}$ model was 5250 K. In our analysis, we infer temperatures of 7500 K in epoch 1 and a drop of 500 K by epoch 3, with a luminosity that increases by a factor of approximately 1.8 between the epochs. This solution roughly matches the $T_\mathrm{eff}$ and luminosity variations of yellow super/hypergiants \citep[e.g.][]{Percy19,vanGenderen19}. However, the required circumstellar extinction is very high ($A_V=2.8$ mag) compared to the majority of such objects, where a total visual dust extinction of $A_V\lesssim2$ is more common \citep[]{Gordon16, Humphreys23}, and the amount of interstellar extinction affecting both components is very low ($A_V=0$ mag in the presented fit, although up to $\approx 0.3$ mag can be accommodated).

Acceptable fits can also be found when only the extinction and effective temperature are kept fixed while allowing only the luminosity to vary between epochs. The increase in luminosity between epochs changes to a factor of approximately 2, which is comparable to the factor obtained when the effective temperature was allowed to vary alongside the luminosity. However, allowing only the effective temperature or only the circumstellar dust extinction to vary between the epochs does not yield any viable solutions.

We conclude that while a solution that includes circumstellar dust extinction can reduce the anomalous $L_\mathrm{hot}/L_\mathrm{cool}$ luminosity ratio, the significant amount of extinction required makes it challenging to reconcile the low-$T_\mathrm{eff}$ star with the super/hypergiants observed in the local Universe.

\section{Scenario II: Two stars from the same star cluster}
\label{Section: Same star cluster}
Two stars do not necessarily need to form a binary system to attain very high magnifications simultaneously. Typically, it is sufficient for them to be projected within $\lesssim 1$ pc of the macrocaustic of a strong-lensing galaxy cluster to reach a total magnification of $\gtrsim 1000$. Considering that only young, high-mass stars at $z\gtrsim 1$ can be rendered detectable by lensing, and given the highly clustered nature of star formation, it seems reasonable to assume that if a highly magnified star is detected, there are likely other high-mass stars in its vicinity. This holds true in cases where the stars are born as members of a bound star cluster, but it may also apply temporarily even if they are formed as part of a dispersing open cluster or OB association. Well-studied young, resolved star clusters like the Orion Nebula Cluster and R136 do indicate an age spread of a few Myr within a 1 pc radius \citep[e.g.][]{Krumholz19,Bestenlehner20}. Whether this age spread is intrinsic to these clusters in three dimensions or is a result of contamination by background/foreground stars within an extended star-forming complex where these clusters reside is not relevant in the current context. What matters instead is the projected age distribution at small separations in the plane of the sky that affects the lensing situation.

\begin{figure}
    \centering
    \includegraphics[width=\columnwidth]{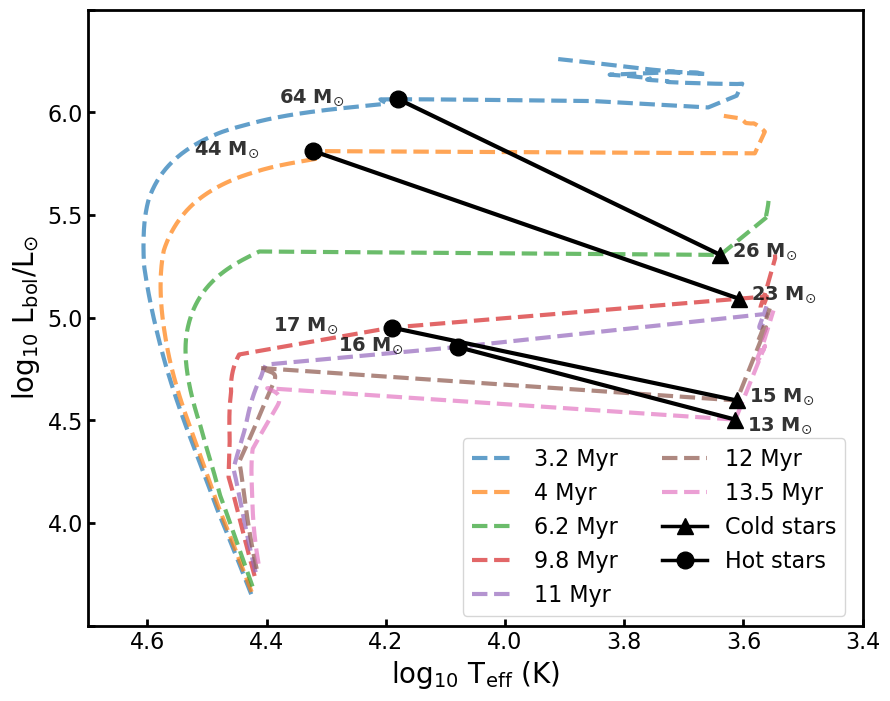}

    \caption{Same as Fig. \ref{fig:solutions_same_age} but for the stars with an age difference of $\leq 3$ Myr. In this case, much larger combinations of stars --within wider age and mass ranges -- meet the $L_\mathrm{hot}/L_\mathrm{cool}>2$ and $T_\mathrm{eff}$ requirements for the two-component SEDs. Here, only a few examples together with their ZAMS masses are shown.}
    \label{fig:solutions_different_ages}
\end{figure}

Hence, we explored whether two stars with ZAMS mass in the range of 9$-498~M_\odot$ at ages up to 27.6 Myr from the same star cluster, with the same metallicity, but with a small age difference may explain the $L_\mathrm{hot}/L_\mathrm{cool}$ conundrum. Considering the previous conditions, we searched for pairs of solutions among different isochrones with an age difference of up to 3 Myr, derived from the \citet{Szecsi22} SMC-metallicity tracks. This resulted in a wider mass ratio range for the pair solutions of $R_{\rm M} = 0.07$--0.89. A few examples of such solutions for various ages and masses are shown in Fig.~\ref{fig:solutions_different_ages}-right. 

To investigate how the age difference between two stars affects the probability of finding such solutions, we performed a simple calculation based on a stellar initial mass function with slope $dN/dM \propto M^{-2.3}$, as suggested by \citet{Kroupa2001IMF} for $M_\mathrm{ZAMS}\geq 1\ M_\odot$. Under the assumption that the age distribution within a star cluster is uniform up to  $\Delta(t)$ = 3 Myr, and that all star-cluster ages are equally likely, we find that the likelihood of encountering these solutions in a star cluster increases by a factor of $\approx 260$, compared to the single-age star cluster case. While this calculation may not accurately capture the likelihood of observing these solutions in samples of lensed stars, which are subject to additional biases related to the magnification distribution, bolometric luminosities, $T_\mathrm{eff}$ and radii of stars \citep{Zackrisson23}, it does lend credence to the idea that stars from the same star cluster -- but with slightly different ages -- offers a viable solution to the $L_\mathrm{hot}/L_\mathrm{cool}$ problem.

\begin{figure*}
\centering
\includegraphics[scale=0.285]{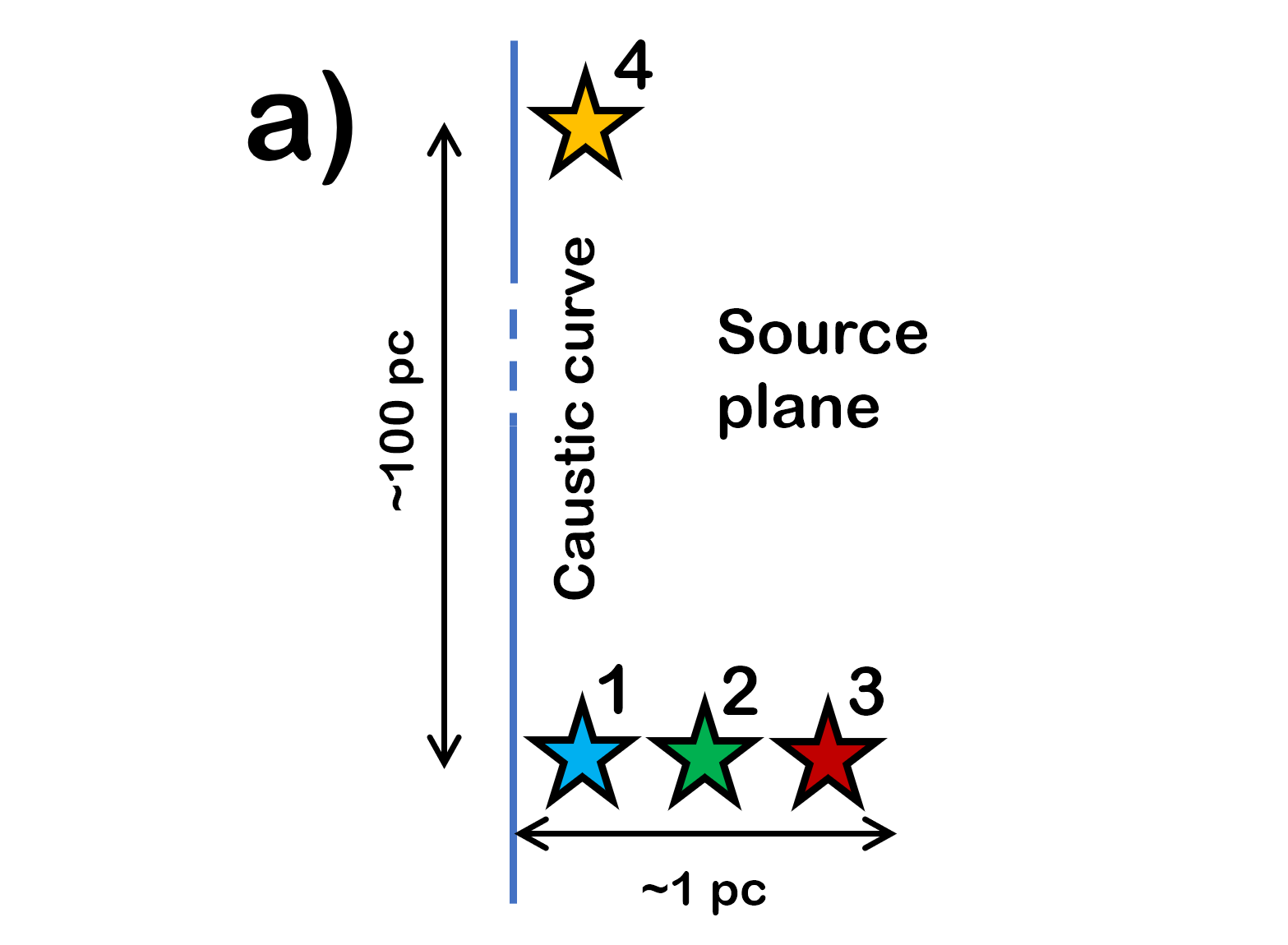}
\includegraphics[scale=0.3]{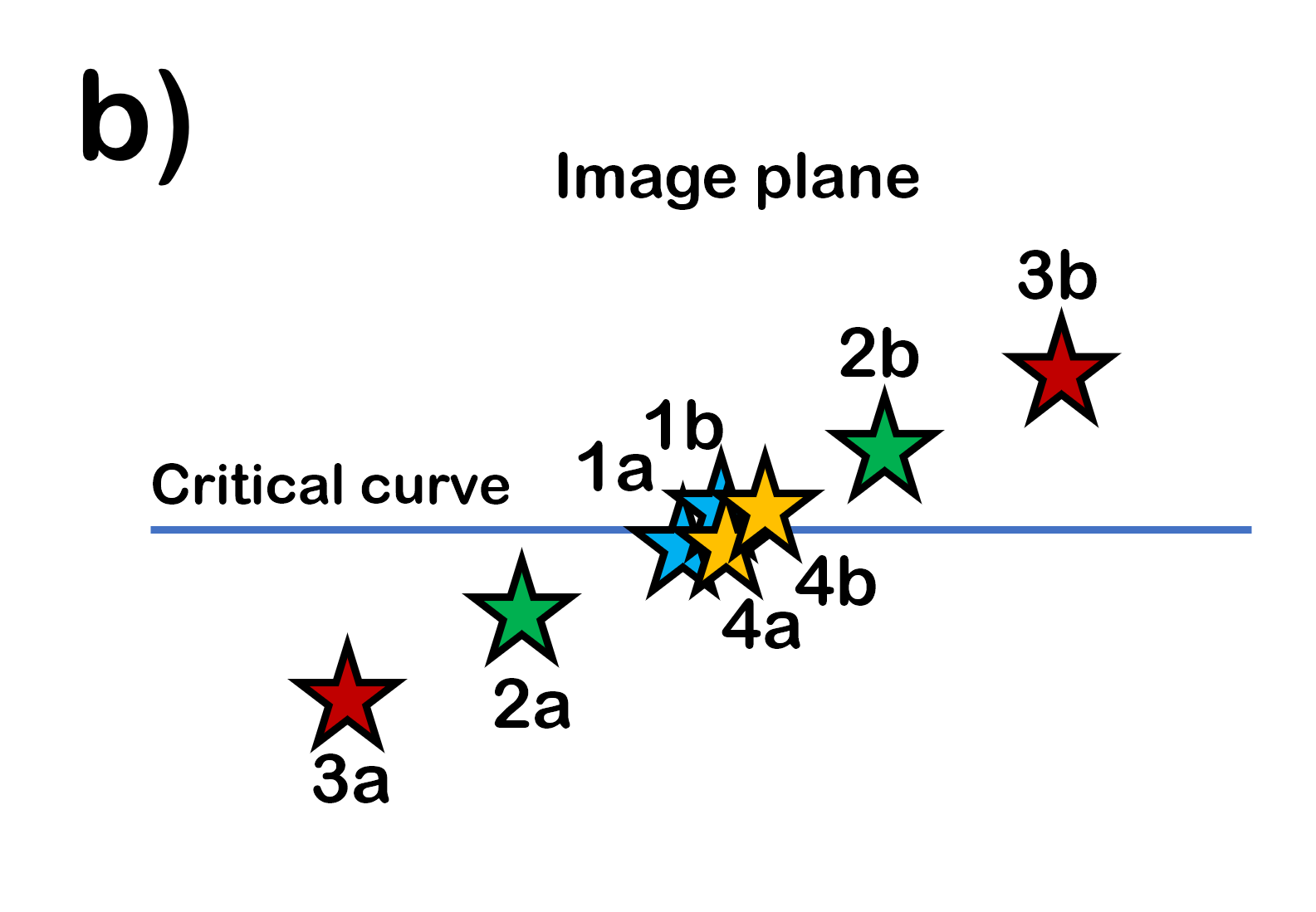}

\caption{Schematic illustration of the lensing situations that allow the light from two stars in separate star clusters to blend together in the image plane forming a two-component SED. {\bf a)} The situation in the source plane, with three stars (1,2, and 3) located close to each other and within $\sim 1$ pc of the caustic curve. A fourth star (4) is located at a similar distance from the caustic as 1, but up to $\approx 100$ pc away in the direction along the caustic. {\bf b)} The corresponding image-plane situation, where the stars 2 and 3 appear as resolved double images (2ab, 3ab) with a separation that increases with their distance from the caustic in a). The images of stars 1 and 4, on the other hand, form two sets of overlapping image pairs (1ab and 4ab) at very high magnification, separated by only a small distance from the critical curve. Since the radial magnification $\mu_\mathrm{r}$ is much smaller than the tangential magnification $\mu_\mathrm{t}$ along the critical curve  (by a factor of $\mu_\mathrm{r}/\mu_\mathrm{t}\lesssim 0.01$), the two image pairs 1ab and 4ab may appear as an unresolved source at the angular resolution of JWST.}
\label{figure: schematic}
\end{figure*}

\section{Scenario III: Two stars from different star clusters}
\label{Section: Different star clusters}

The analysis of the two-component SEDs by \citet{Welch22a,Welch22b,Diego23b,Furtak24} assumes that the two stars would be projected within $<1$ pc of each other in the source plane, but this is not the only possibility. In Figure~\ref{figure: schematic} we schematically illustrate the case where one star is placed at various distances along the direction perpendicular to the macrolensing caustic curve in the source plane. At the largest separation from the caustic, this results in two distinct and well-separated images with intermediate magnification in the image plane. As the distance to the caustic decreases, these images become closer together and more highly magnified and eventually start to blend in the image plane (images 1a and 1b in Figure~\ref{figure: schematic}). However, if a second star is placed similarly close to the caustic, but at a separation of up to $\lesssim 100$ pc along the caustic, a second pair of images (4a and 4b) will appear extremely close to the first pair (1a and 1b), potentially appearing as an unresolved source at the angular resolution of JWST images. This is a consequence of the modest value for the component of the magnification located in the direction along the caustic, $\mu_r$, which typically takes values between 1 and 2, while the component of magnification in the direction perpendicular to the caustic of the magnification, $\mu_t$, can take much larger values, and exceeding 1000 close enough to the caustic. Hence objects near the caustic get amplified by a much larger factor $\mu_t$ in the direction perpendicular to the caustic than in the direction along the caustic, and two objects separated by a large separation in the direction along the caustic can remain unresolved when amplified.

To date, the strongest constraints on the separation between the two stars contributing to SEDs of lensed stars come from \citet{Welch22b} for the $z\approx 6$ star Earendel ($<0.02$ pc). This constraint is based on simulations of the type shown in \citet{Welch22a}, where the separation of two-point sources is increased until the resulting image of stars becomes resolvable and no longer matches the observed point-like morphology. However, these simulations assumed the separation to be in the direction perpendicular to the caustic, rather than along the caustic. By rerunning the simulations, assuming Earendel to cover $\leq 1$ native detector pixel (0.031 arcsec for the NIRCam short-wavelength images) with the separation along the caustic for five different lens models, we obtained the upper limits on the separation presented in Table~\ref{Table:maximum_separation}, which are in the range 85--160 pc.

This demonstrates the viability of solutions to the two-component SED conundrum in which the two massive stars are separated by up to $\approx 100$ pc and hence likely belong to different star clusters and potentially different star-forming regions. In such scenarios, the ages, magnifications and potentially even metallicities of the two stars may differ significantly. Since both stars are likely to have a very high ZAMS mass \citep[$\gtrsim 20\ M_\odot$ in the case of Earendel;][]{Welch22b}, they must both originate from young star clusters. 

\begin{table}[t]
\caption{Maximum separation along the caustic between two point sources contributing to the unresolved JWST morphology of Earendel \citep{Welch22b}.}
\label{Table:maximum_separation}
\centering
\vspace{\fill}
\begin{tabular}{lc}
Lens model & Maximum separation (pc)\\
\hline
LTM           & 107\\
Glafic (c=1)  & 85\\
Glafic (c=7)  & 162\\
WSLAP+        & 155\\
Lenstool      & 160\\
\hline
\end{tabular}
\end{table}


Given that we know there must be at least one such cluster close to the caustic, what is the probability that there is another one within $\approx 100$ pc as well?

\begin{figure*}
\centering 
 \includegraphics[width=8.4cm]{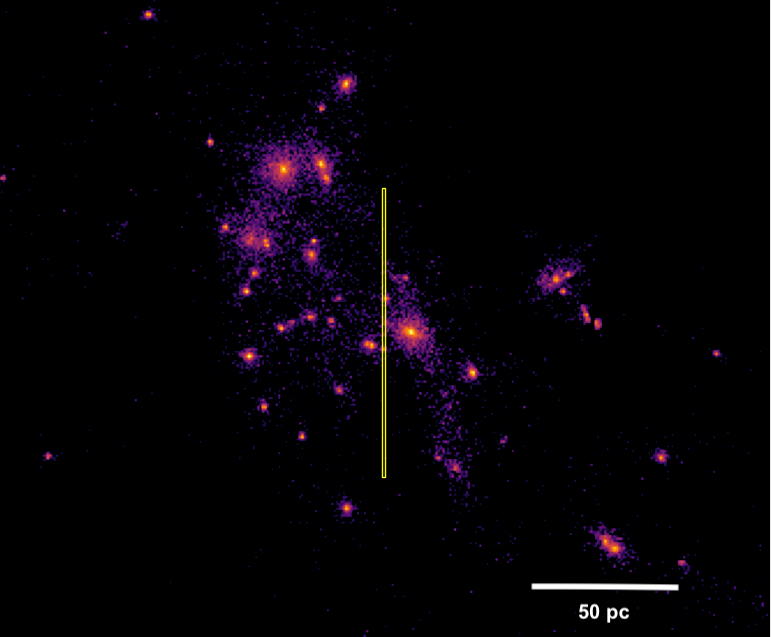}~
 \includegraphics[width=9cm]{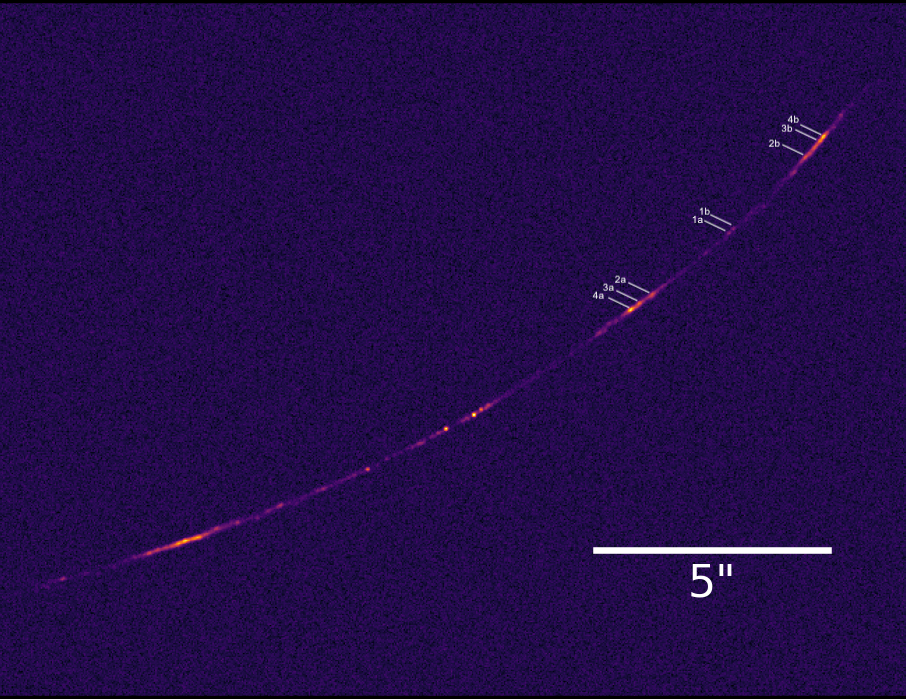}
\caption{Snapshot of a simulated galaxy from \protect{\citet[][]{Garcia2023}} (HSFE run), in which the high star formation efficiency (SFE) in it's star forming gas clouds, which have densities $\sim 100\times$ that of Milky Way's molecular clouds, leads to the formation of numerous compact (radii of $\sim 1-2$ pc) and gravitationally bound star clusters. Here the color coding shows the density of stars and the yellow rectangle of size $1\times 100$~pc represents a sketch of a ray-traced JWST pixel next to the critical curve (line of infinite magnification) traced from the image plane to the the source plane (Left). Lensed image of the galaxy shown on the left panel, placed at $z=6$ across the caustic of the cluster WHL0137-08, using the {\sc glafic} lensing model reproducing the Sunrise Arc/Earendel (Right). See the text for the details of how the image was produced.}
\label{fig:galaxy}
\end{figure*}

In the following, we assess the probability that more than one star-cluster is present within the JWST PSF of a strongly magnified object at $z\sim 6-$10. 
%
%
To quantify this effect, we use a simulation presented in \citet{Garcia2023}, in which the high star formation efficiency (HSFE run) of the star-forming clouds leads to the formation of compact (radii of $\sim 1-$2~pc) and gravitationally bound star clusters. 
The choice is justified by the qualitative agreement between the clumpiness of these simulated high-redshift galaxies \citep[see also,][]{Sugimura2024} and observations of strongly lensed arcs at $z>6$ \citep{Welch22a, Adamo2024, Bradley2024}. This is illustrated in Figure~\ref{fig:galaxy}, showing a snapshot of one of these galaxies (left panel) and the lensed arc it produces (right panel) when using the lens model for the Sunrise Arc/Earendel \citep[][]{Welch22a}. The left panel shows the distribution of the stars in the galaxy, which appear as a collection of bound star clusters, with no prominent bulge or disk. The color coding shows the surface density of the stars projected on the sky; the scale, in parsecs, is shown by the inset white bar, while the yellow rectangle of dimensions 1~pc $\times$100~pc represents the JWST pixel near the critical curve, traced back to the source plane.

The right panel shows the observed flux of the galaxy produced by the stellar continuum at 1500~\AA\ (rest frame), using the {\sc glafic} lens model for the Sunrise Arc \citep{Welch22a} with the simulated galaxy placed at $z=6$ next to the caustic. The scale of the arc, in arcseconds, is indicated by the inset white bar.
Note that the linear scale of the galaxy has been increased by a factor of two to match the size and stellar mass (scaled up by a factor $2^4$) of the $z=6$ galaxy producing the Sunrise Arc. With this scaling, we reproduced the total stellar mass and extent of the arc, as well as its observed flux and clumpiness. Further details on the ray tracing method and the Python package used to create this image will be published in Park et al., in preparation).
In the simulations by \citet{Garcia2023}, star particles have masses of 10~$M_\odot$ and bound star clusters have a power-law mass function with a slope of $\Gamma \sim -1.5 \pm 1$ and masses ranging between 500~$M_\odot$ and $5\times 10^4$~M$_\odot$. A flatter slope, close to $\Gamma =-0.5$ is found in simulations with a more bursty star formation and a clumpy appearance, while a steeper slope is associated with less bursty and more diffuse galaxies. After the scaling mentioned above the masses increase by a factor of about 20 to $10^4$~$M_\odot - 10^6$~M$_\odot$.


The histogram in Fig.~\ref{fig:probability} shows the probability that the light in a strongly magnified pixel near the caustic is produced by a number $N>0$ of star clusters. We used the following procedure to obtain the histogram.
Initially, we used a single snapshot from the simulation (the one shown in Fig.~\ref{fig:galaxy}). We created 50,000 random realizations for the position and rotation of the galaxy relative to the $1\times 100$~pc rectangle. For each realization, we then created a 1D histogram of the star counts falling within it, as a function of the position along its longer side. We smoothed the histogram using a Gaussian kernel with a $\sigma=0.1$~pc and estimated the number of star clusters by counting the number of peaks above a threshold of 4. We also remove stars older than 10~Myr as they would have a negligible contribution to the luminosity and discarded realizations with a total number of stars within the rectangle less than 200, for the same reason. Finally, we produced a histogram of occurrences of $N$ star clusters and normalized it to unity after removing the bin with $N=0$.
We have also done the same calculation including a few hundred snapshots at various cosmic times for the same simulation. However, for the sake of brevity, we do not show this case since the resulting histogram is almost identical to the one shown here.

Fig.~\ref{fig:probability} shows that if enough stars are present near the caustic the probability they are all within one star cluster is $P(1)=25\%$, while the probability of them being in two distinct clusters is higher ($P(2)=35\%$ or $P(2)/P(1)=1.4$). Overall the probability of them being in more than one cluster (i.e., two or more) is $\sim 75\%$, hence very likely.

If we think in terms of a mathematically motivated model for the distribution, assuming that the star clusters are not clustered (i.e., randomly distributed in the sky) we can describe the probability of $N$ star clusters being present in a region of area $A$ within the whole galaxy of area $B$ as Binomial:
\begin{align}
Pr(N) = \binom{n}{N} p^N(1-p)^{(n-N)},
\end{align}
where $n$ is the total number of star clusters in the galaxy and the probability $p=A/B$ is the fraction of the projected area of the galaxy covered by the JWST pixel next to the caustic on the source plane, represented by the yellow rectangle $1~{\rm pc} \times 100~{\rm pc} = 100~{\rm pc}^2$ shown in Figure~\ref{fig:galaxy}-left.
A rough estimate for $p$, accounting for a diameter of the star clusters of 5~pc is $p \approx 1000/10^5 \approx 10^{-2}$. The typical number of star clusters in the galaxy is about 100 (including the low-mass end of the distribution), hence the mean of the distribution is $np \sim 1-2$ and the peak (mode) is $\lfloor (n+1)p\rfloor \approx 1-2$. Since $n$ is reasonably large we can approximate the Binomial with a Poisson distribution: 
\begin{align}
Pr(N) = \frac{\lambda^N e^{-\lambda}}{N!},
\end{align}
where $\lambda=np \sim 1-2$. The probability $P(2)/P(1)=np/2 \sim 0.5-1$. From the simulations we find an even larger ratio, probably because our assumption of a random distribution for the star cluster within the galaxy is incorrect: inspecting Fig.~\ref{fig:galaxy}-left it is evident that star clusters show significant clustering.

Given that it seems quite likely that more than one young, massive star cluster is located close to the caustic, and that the probability for finding solutions that explain the $L_\mathrm{hot}/L_\mathrm{cold}$ problem of lensed stars with two-component SEDs increases when the ages of the stars is allowed to vary (Section~\ref{Section: Same star cluster}), we argue that this scenario (III) is the most likely out of the ones explored.

\section{Discussion and conclusion}
\label{Section: Discussion}

The search for high-redshift lensed stars through HST and JWST observations has revealed a few cases with two-component SEDs, where both a cold and a hot component are involved showing $L_\mathrm{hot}/L_\mathrm{cool}\gtrsim 2$. Given that the two-component SED emerges from a blue supergiant and a yellow/red supergiant, this ratio is unlike that of two stars of the same age, in the context of single-star evolution. 

One may envision scenarios in which the merger of two stars forms a rejuvenated, high-$T_\mathrm{eff}$ and high-$L$ object \citep[e.g.][]{Glebbeek13,Wang22}, which could then be lensed to detectable levels together with an unrelated lower-$L$ star from the same star cluster. If this merger happens in what was originally a triple-star system, the merger product could also potentially find itself bound to a cooler low-$L$ star in a binary. However, if such cases were common, it remains unclear why such stars would not turn up more often in local samples of red supergiant binaries.

Here in this study, we investigated the possible explanations for this luminosity ratio discrepancy by examining three different scenarios:

\begin{figure}
 \centering 
 \includegraphics[width=8cm]{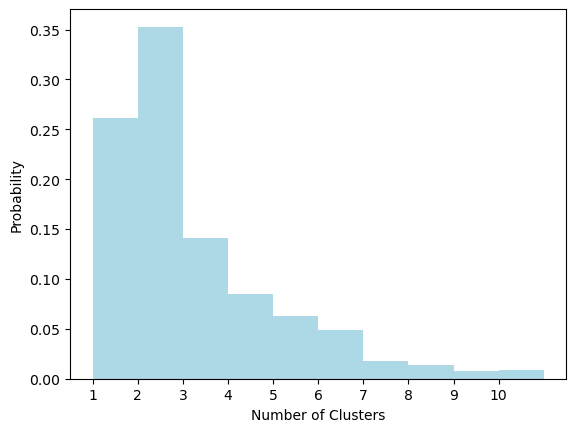}
\caption{Probability distribution (normalized to unity) that $N$ star clusters (excluding the cases with $N=0$) are within 1~pc of the caustic (line of infinite magnification on the source plane) for the different realization of the position and rotation of the galaxy shown in Fig.~1. The histogram shows that if enough stars are present near the caustic the probability they are all in a single star cluster is $P(1)=25\%$, while the probability of them being in two clusters is higher ($P(2)=35\%$ or $P(2)/P(1)=1.4$). The probability of them being in more than one cluster (2 and above) is $\sim 75\%$.
}  \label{fig:probability}
\end{figure}

\begin{itemize}
    \item High intrinsic extinction of the low-$T_{\rm eff}$ component due to circumstellar dust may lower its apparent luminosity, which -- if not corrected for -- leads to an increase in $L_\mathrm{hot}/L_\mathrm{cool}$. However, this would require a substantial amount of such dust around the red components, which are mostly super/hypergiants, a condition not commonly observed in the local Universe.
    \item The observed SED of a magnified object might indeed result from two stars located in the same star cluster, possessing different temperatures and luminosities. However, since stars within a star cluster are generally of similar ages, the likelihood of this scenario decreases as the age difference between the stars becomes smaller.
    \item On the other hand, based on the simulations described above, there is a considerable chance of the stars originating in different star clusters if they are sufficiently close to the caustic curve, with a separation distance of 100 pc along the caustic. Given this scenario, two stars with an age difference $\lesssim 30$ Myr (the maximum lifetime of $\gtrsim 10\Msun$ stars), can become lensed, resulting in an unresolved magnified image where the observed SED would exhibit a dual-component nature.
\end{itemize}

Among the three different scenarios we studied, the third one is more favorable. This scenario, supported by simulations, would explain the two-component spectra of high-z lensed stars, such as Mothra and Erendal. The observed spectra in this scenario, which are a combination of two individual spectra originating from two stars located in different star clusters, could provide a solid explanation for the observed distinct characteristics such as age, luminosity, and temperature. This suggests that the lensed sources might not necessarily be binary systems.

\begin{acknowledgements}
AN and EZ acknowledge funding from Olle Engkvists Stiftelse. EZ also acknowledges grant 2022-03804 from the Swedish Research Council, and has benefited from a sabbatical at the Swedish Collegium for Advanced Study.
\end{acknowledgements}

\bibliographystyle{aa} 
\bibliography{references} 

\end{document}